\documentclass[onecolumn,12pt]{IEEEtran}
\usepackage{amsmath,amssymb}
\usepackage[T1]{fontenc}
\usepackage[dvips]{graphicx, color}
\usepackage{subfigure}
\usepackage{latexsym}
\definecolor{Bgreen}{rgb}{ .0, .55, .0 }
\definecolor{Red}{rgb}{ 1.0, .0, .0 }
\definecolor{Navy}{rgb}{ 0.0, .0, 1.0 }

\newtheorem{e_theo}{Theorem}
\newtheorem{e_defin}{Definition}

\newtheorem{e_prop}{Proposition}

\newtheorem{e_rema}{Remark}

\newcommand{\vect}[1]{\ensuremath{\boldsymbol{#1}}}

\newcommand{\hadjust}[1]{&& \hspace*{#1}}

\newcommand{\eqa}{\begin{eqnarray}}
\newcommand{\ena}{\end{eqnarray}}
\def\rn#1{\expandafter{\romannumeral#1}} 
\def\QED{\hfill$\Box$}

\newcounter{linenumber}
	{\end{list}}

\ifCLASSINFOpdf
\else
\fi
\begin{document}
%
\title{Improved Rate-Equivocation Regions for Secure Cooperative Communication}
%
%
%

\author{Ninoslav~Marina,~\IEEEmembership{Member,~IEEE,}~Hideki~Yagi,~\IEEEmembership{Member,~IEEE,}
        and~H.~Vincent~Poor,~\IEEEmembership{Fellow,~IEEE}
\thanks{N.~Marina is supported by the European Commission under Marie Curie FP7 PEOPLE Programme, Grant \#237669.  H.~Yagi is supported in part by MEXT under  Grant-in-Aid for Young Scientists (B) No.\ 22760270, JST's Special Coordination Funds for Promoting Science and Technology, and H.~V.~Poor is supported by the U.S.\ National Science Foundation under Grant CNS-09-05398.}
\thanks{N.\ Marina is with Department of Electrical Engineering, Princeton University, Princeton, NJ 08544 USA (email: nmarina@princeton.edu).}
\thanks{H.\ Yagi is with Center for Frontier Science and Engineering, The University of Electro-Communications, Chofu-shi, Tokyo 182-8585, Japan (email: yagi@ice.uec.ac.jp).}
\thanks{H.\ V.\ Poor is with Department of Electrical Engineering, Princeton University, Princeton, NJ 08544 USA (email: poor@princeton.edu).}
\vspace{-.5cm}
}
\maketitle

\begin{abstract}
A simple four node network in which cooperation improves the information-theoretic secrecy is studied. The channel consists of two senders, a receiver, and an eavesdropper. One or both senders transmit confidential messages to the receiver, while the eavesdropper tries to decode the transmitted message. The main result is the derivation of a newly achievable rate-equivocation region that is shown to be larger than a rate-equivocation region derived by Lai and El Gamal for the relay-eavesdropper channel. When the rate of the helping interferer is zero, the new rate-equivocation region reduces to the capacity-equivocation region
over the wire-tap channel, hence, the new achievability scheme can be seen as a generalization of a coding scheme proposed by Csisz\'ar and K\"orner. This result can naturally be combined with a rate-equivocation region given by Tang
et al. (for the interference assisted secret communication), yielding an even larger
achievable rate-equivocation region. 
%
\end{abstract}


\begin{IEEEkeywords}Information-theoretic secrecy, wire-tap channel, eavesdropper channel, rate-equivocation region, secrecy capacity, perfect secrecy, physical layer security, cooperative communication.
\end{IEEEkeywords}

%
\IEEEpeerreviewmaketitle

\section{Introduction}\label{intro}
%
%
%
%
\IEEEPARstart{I}n this work we propose a scheme that increases the information theoretic secrecy in a simple cooperative communication network. The channel model includes a class of the wire-tap channels with a helping interferer introduced by Lai and El Gamal \cite{Lai-ElGamal2008}. These authors considered several cooperation schemes over the relay-eavesdropper channel, in which the relay node helps to enhance the security level of communication between the sender and the receiver. The paper gives an interesting observation indicating that over the multiple access channel (MAC) with an eavesdropper, secret communication can be enhanced with a help of one of the two senders (called, the \emph{helping interferer} or the \emph{helper}). In addition, an achievable equivocation-rate region has been derived for this scheme. Subsequently, Tang et al.~\cite{TLPP2008} have derived an improved rate-equivocation region using the fact that the receiver does not have to decode the sequence transmitted by a helper. One possibility is that the helper sends interference (dummy messages) in order to weaken the channel to the eavesdropper. When the rate of dummy messages of the helper is zero (there is no cooperation from the helper), the channel reduces to the (single-user) \emph{wire-tap channel} introduced by Wyner \cite{Wyner75}, and generalized later by Csisz\'ar and K\"orner \cite{Csiszar-Korner78}. In this reduced setting, however, the achievable rate-equivocation regions given by \cite{Lai-ElGamal2008} and \cite{TLPP2008} do not coincide with the capacity-equivocation region over the wire-tap channel, giving only its sub-region. 
 When only perfect-secrecy is imposed (i.e., the eavesdropper is totally ignorant of the transmitted message), their results coincide with the secrecy-capacity of the wire-tap channel.

Motivated by this fact, the first part of this paper gives a new achievable rate-equivocation region (i.e, an inner bound on the capacity-equivocation region) for the wire-tap channel with a helping interferer, showing that the new region is improved over the one given in \cite{Lai-ElGamal2008}. When the rate of the helping interferer is zero, the new rate-equivocation region reduces to the capacity-equivocation region over the wire-tap channel, so the new achievability scheme can be seen as a generalization of the coding scheme given in \cite{Csiszar-Korner78}. Our result can naturally be combined with the additional rate-equivocation region given by \cite{TLPP2008}, yielding an even larger rate-equivocation region.

In the next section we present the previous results on the wire-tap channel with a helper. The main result of this work, that is the improved rate-equivocation region for the wire-tap channel with helping interferer, is presented in Section~\ref{sect:new_RE_region}, while in Section~\ref{sect:achievability_R1} we derive an even larger rate-equivocation region. A note on the broadcast channel with confidential messages and the wire-tap channel with helping interferer is given in Section V. Section~\ref{sect:conclusion} concludes the paper.



\section{Preliminaries} \label{sec:system_model}
\subsection{The Wire-Tap Channel with a Helping Interferer}

\begin{figure}[h]
\begin{center}
\input{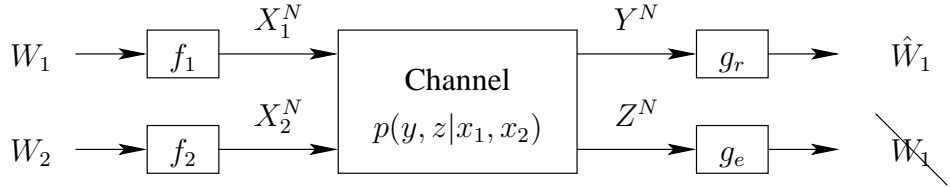}
\end{center}
\caption{A four node network of one sender, one receiver, one eavesdropper and one helper.}
\label{fig1}
\end{figure}

The cooperative channel considered in this paper is shown in Fig.~\ref{fig1} and consists of two senders, a receiver, and an eavesdropper, in which one sender transmits confidential messages to the receiver, and the eavesdropper tries to decode the transmitted message.
The second sender plays the role of a ``helper'' to enhance the secrecy of communication. This model is referred to as the wire-tap channel with a (helping) interferer, and will be considered first.
Let $\mathcal{X}_t$ be the channel input alphabet of sender $t$, $t=1,2$, and let and $\mathcal{Y}$ and $\mathcal{Z}$ be the output alphabet of the receiver and the eavesdropper, respectively. We assume that all the alphabets are discrete and finite and the channel is memoryless, characterized by a conditional probability mass function (PMF) $P(y,z|x_1x_2)$ for $(x_1, x_2) \in \mathcal{X}_1 \times \mathcal{X}_2$ and $(y, z) \in \mathcal{Y} \times \mathcal{Z}$,
{i.e.,} $\vect{x}_t\triangleq(x_{t1}, \ldots, x_{tN}) \in \mathcal{X}_t^N$, $\vect{y}\triangleq(y_1, \ldots, y_N) \in \mathcal{Y}^N$ and $\vect{z}\triangleq(z_1, \ldots, y_N) \in \mathcal{Y}^N$. Then, we have
\newcommand{\n}{\nonumber}
\begin{equation} \label{eq:memoryless}
P_N(\vect{y},\vect{z}|\vect{x}_1,\vect{x}_2) = \prod_{n = 1}^N P(y_{n},z_{n}|x_{1n},x_{2n})\n
\end{equation}
where $N$ denotes the number of channel uses. We assume that both of the receiver and the eavesdropper know $P(y,z|x_1,x_2)$. 

Define $\mathcal{W}_t$ with $t =1,2$ as the set of integers $\{1, \ldots, M_t\}$ with $M_t \ge 1$.  Let $w_1 \in \mathcal{W}_1$ be a uniformly distributed confidential message of sender 1. We also denote a random message of sender 2 by $w_2 \in \mathcal{W}_2$. Encoder $t$ is a deterministic mapping denoted by
\begin{equation}
f_t : \mathcal{W}_t \rightarrow \mathcal{X}_t^N.
\end{equation} 
The receiver and the eavesdropper estimate the transmitted message from the received sequence $\vect{y}$ and $\vect{z}$, with the decoding functions
\begin{equation}
g_r : \mathcal{Y}^N \rightarrow \mathcal{W}_1, ~~\mbox{and}~~
g_e : \mathcal{Z}^N \rightarrow \mathcal{W}_1,\n
\end{equation}
respectively. Let $R_t$, $t =1,2$, be an information rate defined as
\begin{equation}
R_t = {\log_2 M_t}/{N}. \n
\end{equation}
An $(N,M_1,M_2,\{f_t,g_t\}_{t=1,2})$ code for the MAC with a helper consists of message sets $\mathcal{V}_1 \times \mathcal{V}_2$, encoding functions $f_t$, and decoding functions $g_t$ with $t=1,2$. Provided that the transmitted message is $w_1\in \mathcal{W}_1$, the decoder makes an \emph{error} if $g_r(\vect{y}) \neq w_1$. The average probability of decoding error, denoted by $P_{\rm e}^{(N)}$, is
\begin{equation}
P_{\rm e}^{(N)} = \frac{1}{M_1} \sum_{w_1 \in \mathcal{W}_1} \Pr(g_r(\vect{y}) \neq w_1 |~w_1 ~\mbox{sent}).\n
\end{equation}
The \emph{equivocation} rate at the eavesdropper is defined as
\begin{equation}
R_{\rm e}^{(N)} = \frac{1}{N} H(W_1|Z^N).\n
\end{equation}
The secrecy considered in this paper is defined as follows:
\begin{e_defin}\label{def:R-E pair}
A \emph{rate-equivocation pair} $(R_1, R_{\rm e})$ is said to be \emph{achievable} if there exists a sequence of $(N,M_1)$ codes such that for every $\epsilon > 0$,
\begin{eqnarray}
R_1  \ge  \frac{\log_2 M_1}{N} - \epsilon,  \quad P_{\rm e}^{(N)} \le \epsilon,  \quad\mbox{and}\quad R_{\rm e}^{(N)} \ge  R_{\rm e} -\epsilon,\n
\end{eqnarray}
for all sufficiently large $N$.
\end{e_defin}

\begin{e_defin}
A \emph{perfect-secrecy rate} $R_1$ is said to be \emph{achievable} if the rate-equivocation pair $(R_1, R_1)$ is achievable.
The \emph{secrecy-capacity} of the wire-tap channel with a helper is defined as the maximum of all achievable perfect-secrecy rates.
\end{e_defin}

Note that without sender 2, this channel model reduces the (single-user) wire-tap channel \cite{Wyner75,Csiszar-Korner78}.
Achievable rate-equivocation pairs, achievable perfect-secrecy rates, and the secrecy capacity for the wire-tap channel are defined analogously.

\subsection{Known Achievable Rate-Equivocation Regions}
For the (single-user) wire-tap channel \cite{Wyner75,Csiszar-Korner78}, the following rate-equivocation region is the capacity-equivocation region
\begin{eqnarray}
\bigcup_{P_{QU}P_{X_1|U} P_{YZ|X}} \Big\{ (R_1, R_{\rm e}): &&0 \le R_{\rm e} \le R_1 , \n\\
&&R_1 \le I(U;Y), \nonumber \\
&&R_{\rm e} \le I(U;Y|Q) -I(U;Z|Q) \Big\}, \label{eq:wiretap_CE_region}
\end{eqnarray}
where $Q$ and $U$ are auxiliary random variables satisfying the Markov chain condition \[Q \rightarrow U \rightarrow X_1\] and the cardinality bounds \[|\mathcal{Q}| \le |\mathcal{X}_1| + 3 \quad \mbox{and} \quad |\mathcal{U}| \le |\mathcal{X}_1|^2 + 4 |\mathcal{X}_1| + 3.\] 
 {Since we assume that all rates in this paper are always non-negative, if an upper-bound on $R_{\rm e} $ happens to be negative, it means $R_{\rm e}=0$. This rule will be applied throughout the paper when necessary.}

For the wire-tap channel with a helper, it was shown in \cite[Theorem 3]{Lai-ElGamal2008} that the following rate-equivocation region is achievable:  
\begin{eqnarray}
\mbox{co}\bigcup_{P_{U_1}P_{U_2} P_{X_1|U_1} P_{X_2|U_2} P_{YZ|X_1X_2}} \Big\{ (R_1, R_{\rm e}):  R_1 &\le& I(U_1;Y|U_2), \n\\
0&\le& R_{\rm e} \le R_1, \nonumber \\
R_{\rm e} &\le& I(U_1;Y|U_2) - \min\{I(U_2;Y), I(U_2;Z)\} \n\\
&-& I(U_1;Z|U_2) + \min\{I(U_2;Y), I(U_2;Z|U_1) \} \Big\}, \label{eq:achievable_region1}
\end{eqnarray}
where 
co$(\mathcal{S})$ denotes the convex hull of the set $\mathcal{S}$,  $U_1$ and $U_2$ are auxiliary random variables satisfying the Markov chain condition \[(U_1,U_2)  \rightarrow (X_1, X_2) \rightarrow (Y,Z).\]

 We note that if $I(U_2;Y) \le I(U_2;Z)$, then the last inequality on $R_{\rm e}$ becomes \[0 \le R_{\rm e} \le I(U_1;Y|U_2) - I(U_1;Z|U_2),\] implying that the wire-tap channel with a helper becomes the ordinary wire-tap channel, i.e., there is no effect from the user cooperation. In this case, the region given by (2) reduces to a sub-region of the region given by (\ref{eq:wiretap_CE_region}). {Note that the result of Tang et al.\ \cite{TLPP2008} implies that we might still have an advantage from the user cooperation in this case.}

For the wire-tap channel with a helper, it is known that the following perfect-secrecy rate is achievable \cite[eq.\ (10)]{Lai-ElGamal2008}: 
\begin{eqnarray}
R_1 = \sup_{P_{U_1}P_{U_2} P_{X_1|U_1} P_{X_2|U_2}}  \big[I(U_1;Y|U_2) - I(U_1;Z|U_2) 
&+& \min\{I(U_2;Y), I(U_2;Z|U_1) \}\n\\
&-& \min\{I(U_2;Y), I(U_2;Z)\}\big]^+, \label{eq:achievable_PS_rate}
\end{eqnarray}
where $[x]^+$ denotes $\max\{x,0\}$. 

\section{Improved Rate-Equivocation Region}
\label{sect:new_RE_region}

In this section we show that it is possible to have a rate-equivocation region larger than the one given by (\ref{eq:achievable_region1}). To that end we introduce an
auxiliary random variable $Q_1$ and we get the following improved region.  
\begin{e_prop} The following rate-equivocation region is achievable:
\begin{eqnarray}
\mathcal{C} = \bigcup_{\pi} \Big\{ &&(R_1, R_{\rm e}): 0 \le R_{\rm e} \le R_1, \nonumber \\
&&R_1 \le R_1' + \min\{ I(\textcolor{black}{Q_1};Y|U_2), I(\textcolor{black}{Q_1};Z|U_2)\}, \nonumber \\
&&R_{\rm e} \le \max \Big\{ R_1' + R_2' - I(U_1;Z|U_2\textcolor{black}{Q_1}) -I(U_2;Y|\textcolor{black}{Q_1}), 
R_1' + R_2' - I(U_1U_2;Z|\textcolor{black}{Q_1}) \Big\} \label{eq:achievable_region2}
\end{eqnarray}
where 
\eqa
\pi&\triangleq& P_{\textcolor{black}{Q_1}}  P_{U_1|\textcolor{black}{Q_1}}P_{U_2} P_{X_1|U_1} P_{X_2|U_2} P_{YZ|X_1X_2}, \n\\
R_1'&\triangleq& I(U_1;Y|U_2\textcolor{black}{Q_1}), \mbox{and} \n\\
R_2'&\triangleq& \min\{I(U_2;Y|\textcolor{black}{Q_1}), I(U_2,Z|U_1) \}.\n
\ena 
$Q_1$, $U_1$, and $U_2$ are auxiliary random variables satisfying the following Markov chain conditions:
\begin{eqnarray}
Q_1 &\rightarrow& U_1  \quad\rightarrow\quad X_1, ~\mbox{and} \nonumber \\
(U_1, U_2) &\rightarrow& (X_1, X_2) \quad\rightarrow\quad (Y,Z). \label{eq:Markov_chain1}
\end{eqnarray}
\end{e_prop}

As in \cite{Csiszar-Korner78}, let the auxiliary random variable $Q_1$ correspond to the sequence alphabet decoded by both the receiver and the eavesdropper, while letting $U_1$ and $U_2$ denote the sequence alphabets that can be decoded only by the receiver. First, we note that the constraint on $R_{\rm e}$ in (\ref{eq:achievable_region2}) can be re-written as
\begin{eqnarray}
\hadjust{-7mm}  
R_{\rm e} \le 
\begin{cases}
I(U_1;Y|U_2\textcolor{black}{Q_1}) - I(U_1;Z|U_2\textcolor{black}{Q_1}), &\mbox{if}~~I(U_2;Y|\textcolor{black}{Q_1}) \le I(U_2;Z|\textcolor{black}{Q_1}),  \\
I(U_1U_2;Y|\textcolor{black}{Q_1}) - I(U_1U_2;Z|\textcolor{black}{Q_1}), &\mbox{if}~~I(U_2;Z|\textcolor{black}{Q_1}) \le I(U_2;Y|\textcolor{black}{Q_1}), \le I(U_2;Z|U_1), \\
I(U_1;Y|U_2\textcolor{black}{Q_1}) - I(U_1;Z|\textcolor{black}{Q_1}), &\mbox{if}~~I(U_2;Z|U_1)  \le I(U_2;Y|\textcolor{black}{Q_1}).\n
\end{cases}
\label{eq:another_expression}
\end{eqnarray}
It is straightforward that by setting $\mathcal{Q}_1 = \emptyset$, we have
\begin{eqnarray}
&&I(U_1;Y|U_2\textcolor{black}{Q_1}) + \min\{ I(\textcolor{black}{Q_1};Y|U_2), I(\textcolor{black}{Q_1};Z|U_2)\} = I(U_1;Y|U_2),\n\\
&&I(U_1;Y|U_2\textcolor{black}{Q_1}) - I(U_1;Z|U_2\textcolor{black}{Q_1}) = I(U_1;Y|U_2) - I(U_1;Z|U_2).\n
\end{eqnarray}
On the other hand, since
\begin{eqnarray}
I(U_1;Y|U_2\textcolor{black}{Q_1}) - I(U_1;Z|U_2\textcolor{black}{Q_1}) = I(U_1;Y|U_2) - I(U_1;Z|U_2) + \big( I(\textcolor{black}{Q_1};Z|U_2) - I(\textcolor{black}{Q}_1;Y|U_2) \big),\n
\end{eqnarray}
we have
\begin{eqnarray}
\sup_{P_{Q_1}P_{U_1|Q_1}P_{U_2}} \big\{ I(U_1;Y|U_2\textcolor{black}{Q_1}) - I(U_1;Z|U_2\textcolor{black}{Q_1}) \big\} \ge \sup_{P_{U_1}P_{U_2}} \big\{ I(U_1;Y|U_2) - I(U_1;Z|U_2) \big\}. \label{eq:Q1_relation1}
\end{eqnarray}
A similar derivation of (\ref{eq:Q1_relation1}) yields
\begin{eqnarray}
  \sup_{P_{Q_1}P_{U_1|Q_1}P_{U_2}} \big\{ I(U_1U_2;Y|\textcolor{black}{Q_1}) - I(U_1U_2;Z|\textcolor{black}{Q_1}) \big\} \ge \sup_{P_{U_1}P_{U_2}} \big\{ I(U_1U_2;Y) - I(U_1U_2;Z) \big\}, \n\label{eq:Q1_relation2}
\end{eqnarray}
and
\begin{eqnarray}
\sup_{P_{Q_1}P_{U_1|Q_1}P_{U_2}} \big\{ I(U_1;Y|U_2\textcolor{black}{Q_1}) - I(U_1;Z|\textcolor{black}{Q_1}) \big\}  \ge \sup_{P_{U_1}P_{U_2}} \big\{ I(U_1;Y|U_2) - I(U_1;Z) \big\}. \n\label{eq:Q1_relation3}
\end{eqnarray}
Hence, region $\mathcal{C}$ given by (\ref{eq:achievable_region2}) is larger than or equal to the region given by (\ref{eq:achievable_region1}). The random variable $ Q_1$ plays not only the role of convexification.
 The achievability of the region $\mathcal{C}$ will be shown in Appendix~\ref{Append:achievability}.

For the rate-equivocation region $\mathcal{C}$, if $I(U_2;Y|Q_1) \le I(U_2;Z|Q_1)$ for every \[P_{Q_1U_1U_2X_1X_2}^*\triangleq P_{Q_1U_1}P_{X_1|U_1}P_{U_2X_2},\] the cooperation between sender 1 and sender 2 (the helper) has no effect, and the region is in a simpler form as the convex hull of 
\begin{eqnarray}
\tilde{\mathcal{C}} = \bigcup_{P_{\textcolor{black}{Q_1}U_1U_2X_1X_2}^* P_{YZ|X_1X_2}} \Big\{ (R_1, R_{\rm e}):  &&0 \le R_{\rm e} \le R_1,  \nonumber \\
&&R_1 \le I(U_1;Y|U_2) \nonumber \\
&&R_{\rm e} \le I(U_1;Y|U_2\textcolor{black}{Q_1}) - I(U_1;Z|U_2\textcolor{black}{Q_1}) \Big\}. \n\label{eq:achievable_region3}
\end{eqnarray}
{Although Tang et al.\ \cite{TLPP2008} give a larger region in this case, if $I(U_2;Y|U_1) \ge I(U_2;Z|U_1)$, then user cooperation does not take effect.} In the following text, we denote the convex hull of $\mathcal{C}$ and $\tilde{\mathcal{C}}$ by $\mathcal{C}^*$ and  $\tilde{\mathcal{C}}^*$, respectively. When there is no helping interference, i.e., $R_2 =0$, then the region $\tilde{\mathcal{C}}$ corresponds to the capacity-equivocation region for the ordinary wire-tap channel given by (\ref{eq:wiretap_CE_region}). Note that in the case $R_2 =0$, the helper transmits a deterministic sequence $u_2^N \in \mathcal{U}_2^N$, and both the receiver and the eavesdropper know this sequence. Therefore, the capacity-equivocation region is still characterized by $U_2$.

When considering the perfect-secrecy rate, the auxiliary random variable $Q_1$ introduced to derive a new rate-equivocation region has no impact. For the wire-tap channel with a helper, we can achieve the same perfect-secrecy rate as (\ref{eq:achievable_PS_rate}) derived in \cite{Lai-ElGamal2008}.

As the last result of this section we get the following theorem.
\begin{e_theo} \label{theo:achievability_R1}
The rate-equivocation region $\mathcal{C}^*$ is achievable for the wire-tap channel with a helping interferer.
\end{e_theo}
{\it Proof:}
From the above argument, by the coding scheme given in Appendix~\ref{Append:achievability}, the region $\mathcal{R}_1(P_{Q_1X_1X_2}^*)$ is achievable for any given $P_{Q_1X_1X_2}^*$, and hence $\mathcal{R}_1$ is achievable. By prefixing a conditional PMF $P_{X_1|U_1}P_{X_2|U_2}$, the region $\mathcal{R}_2$, which is equivalent to $\mathcal{C}$, is also achievable. The convex hull can be taken since we can time-share multiple input PMFs via the \emph{time-sharing principle} \cite{Cover-Thomas91}.

\QED


\section{An Even Larger Rate-Equivocation Region} \label{sect:achievability_R1}

We can combine the idea given in \cite{TLPP2008} with the achievable region $\mathcal{C}^*$ to get a larger achievable rate-equivocation region. The key observation is that the receiver does not necessarily need to decode the dummy message $W_2$ sent from the helper.

For a fixed $P_{Q_1U_1U_2X_1X_2}^* \triangleq P_{Q_1U_1} P_{X_1|U_1} P_{U_2X_2} \in \mathcal{P}^*$, let $\mathcal{C}_A(P_{Q_1U_1U_2X_1X_2}^*)$ be defined as the rate-equivocation region
\begin{eqnarray}
\mathcal{C}_A(P_{Q_1U_1U_2X_1X_2}^*) =  \Big\{&&(R_1, R_{\rm e}): 0 \le R_{\rm e} \le R_1, \nonumber \\
&&R_1 \le I(U_1;Y|U_2Q_1) + \min\{ I(Q_1;Y|U_2), I(Q_1;Z|U_2)\}, \nonumber \\
&&R_{\rm e} \le \max \Big\{ R_3' - I(U_1;Z|U_2Q_1) - I(U_2;Y|Q_1), R_3' - I(U_1U_2;Z|Q_1) \Big\} \label{eq:achievable_region5}
\end{eqnarray}
where 
\eqa
R_2'&=&\min\{I(U_2;Y|Q_1), I(U_2,Z|U_1) \},\n\\ 
R_3'&\triangleq& I(U_1;Y|U_2Q_1)+ R_2'\n,
\ena
and $Q_1$, $U_1$, and $U_2$ are auxiliary random variables satisfying Markov chain conditions (\ref{eq:Markov_chain1}). Then the achievable rate-equivocation region $\mathcal{C}$ is expressed as
\begin{equation}
\mathcal{C} = \bigcup_{P_{Q_1U_1U_2X_1X_2}^* P_{YZ|X_1X_2}} \mathcal{C}_A (P_{Q_1U_1U_2X_1X_2}^*). \label{eq:union_region1}
\end{equation}
We define another rate-equivocation region, for a fixed $P_{Q_1U_1U_2X_1X_2}^* \in \mathcal{P}^*$, as
\begin{eqnarray}
\mathcal{C}_B(P_{Q_1U_1U_2X_1X_2}^*) =  \Big\{ (R_1, R_{\rm e}): &&R_1 \le I(U_1;Y), \nonumber \\
&& 0 \le R_{\rm e} \le R_1, \nonumber \\
&& R_{\rm e} \le  I(U_1;Y|Q_1) - I(U_1;Z|Q_1) \Big\}. \label{eq:achievable_region5b}\n
\end{eqnarray}
Then, a new achievable rate-equivocation region, denoted by $\tilde{\mathcal{C}}$, is given by the convex hull of
\begin{equation}
\tilde{\mathcal{C}} = \bigcup_{P_{Q_1U_1U_2X_1X_2}^* } \big\{ \mathcal{C}_A (P_{Q_1U_1U_2X_1X_2}^*) \cup \mathcal{C}_B (P_{Q_1U_1U_2X_1X_2}^*) \big\}. \label{eq:union_region2}
\end{equation}
From equations (\ref{eq:union_region1}) and (\ref{eq:union_region2}), it is readily seen that in general we have $\mathcal{C}^* \subseteq \tilde{\mathcal{C}}^* $ where $\tilde{\mathcal{C}}^*$ denotes the convex hull of $\tilde{\mathcal{C}}$.
The region \[\mathcal{C}_B(P^*) \setminus (\mathcal{C}_A(P^*) \cap \mathcal{C}_B(P^*))\] expresses an additional region to $\mathcal{C}_A(P^*)$ for a fixed $P^* \in \mathcal{P}^*$, which is  given by the observation in \cite{TLPP2008}.
The rate-equivocation region $\tilde{\mathcal{C}}$ can be seen as an extension of the result of \cite{TLPP2008} in the sense that we derive not only a perfect-secrecy rate but also a rate-equivocation region by introducing the auxiliary random variable $Q_1$.
The key idea lies in the facts that:
\begin{itemize}
\item[(\rn{1})] The receiver and the eavesdropper can decode a partial message of $W_1$ at the rate at most \[\min\{ I(Q_1;Y), I(Q_1;Z) \}, \quad\mbox{and},\]
\item[(\rn{2})] As for the other part of message, dummy message from the helper needs not be decoded, and can be treated as noise.
\end{itemize}
Note that even though the region $\mathcal{C}_B(P_{Q_1U_1U_2X_1X_2})$ does not involve the rate $R_2$, user cooperation, i.e., interference by a helper, is necessary to achieve this region, and hence, the PMFs of random variables $U_2$ and $X_2$ are also included in the region. The achievability of the region $\tilde{\mathcal{C}}$ is shown in Appendix\ \ref{append:RB}.

\begin{figure}[t]
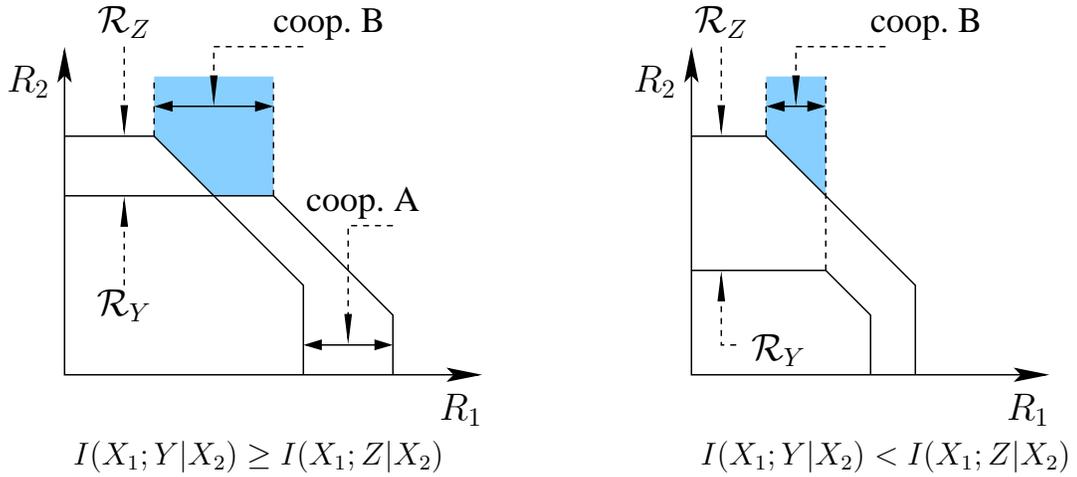

\begin{center}
\input regions_ab_1column.pstex_t
\caption{Pictorial representation for the equivocation gain when cooperation is used for the case (\rn{1}) of Proposition~\ref{TangProp} for the situations $I(X_1;Y|X_2)\ge I(X_1;Z|X_2)$ (left) and $I(X_1;Y|X_2)\ge I(X_1;Z|X_2)$ (right). Pentagons $\mathcal{R}_Y$ and $\mathcal{R}_Z$ express an achievable region for the receiver's MAC and the eavesdropper's MAC, respectively. The cooperation scheme that achieves $\mathcal{C}_A(P^*)$ is labeled ``coop. A'' and the cooperation scheme that achieves $\mathcal{C}_B(P^*)$, is labeled ``coop. B''.}
\label{fig:region_CB1}
\end{center}
\end{figure}

When only the perfect-secrecy rate is concerned, the obtained rate-equivocation region is reduced to
\begin{equation}
\tilde{\mathcal{C}}' = \bigcup_{\pi_{12}} \big\{ \mathcal{C}_A' (\pi_{12}) \cup \mathcal{C}_B' (\pi_{12}) \big\}, \label{eq:union_region3}
\end{equation}
where 
\begin{eqnarray}
\mathcal{C}_A'(\pi_{12}) \triangleq  \Big\{ R_1: &&R_1\ge 0,\n\\
&&R_1 \le \max \big\{ I(U_1;Y|U_2) - I(U_1;Z|U_2) + R_2' -I(U_2;Y), \n\\
&&\hspace{22mm}I(U_1;Y|U_2) + R_2' - I(U_1U_2;Z) \big\} \Big\} \label{eq:achievable_region5c}\n
\end{eqnarray}
and
\begin{eqnarray}
\hadjust{-5mm}  \mathcal{C}_B'(\pi_{12}) =  \Big\{ R_1: 0\le R_1 \le  [I(U_1;Y) - I(U_1;Z)]^+ \Big\}. \n \label{eq:achievable_region5d}
\end{eqnarray}
for a fixed input distribution $\pi_{12}=P_{U_1X_1}P_{U_2X_2}$. Then, the following perfect-secrecy rate is achievable:
\begin{eqnarray}
\hadjust{-7mm}\sup_{\pi_{12}} \big\{\mathcal{C}_A' (\pi_{12}) \cup \mathcal{C}_B' (\pi_{12})  \big\}\n
\end{eqnarray}
 which is the same as the one given in \cite{TLPP2008}.

We next consider conditions under which we get an improvement to region $\mathcal{C}_B(P^*)$, i.e., 
\[\mathcal{C}_B(P^*) \setminus (\mathcal{C}_A(P^*) \cap \mathcal{C}_B(P^*)) \neq \emptyset.\] 
We have the following proposition:

 \begin{e_prop}\label{TangProp}
 For a given $P^* \in \mathcal{P}^*$, $\mathcal{C}_B(P^*) \setminus (\mathcal{C}_A(P^*) \cap \mathcal{C}_B(P^*)) \neq \emptyset$ if and only if either of the following two conditions is satisfied:
\eqa
&&(\rn{1}) \quad I(U_1;Y|Q) > I(U_1;Z|Q) \quad\mbox{and}\quad\n\\
 && \hspace{10mm}0 \le I(U_2;Z|Q_1) - I(U_2;Y|Q_1) \le I(U_2;Z|U_1) - I(U_2;Y|U_1),  \label{eq:effective_condition1}\\
&&(\rn{2})\quad I(U_1;Y|Q) > I(U_1;Z|Q) \quad\mbox{and}\quad\n\\
 &&\hspace{10mm}I(U_2;Z|Q_1) \le I(U_2;Y|Q_1) \le I(U_2;Y|U_1) \le I(U_2;Z|U_1).  \label{eq:effective_condition2}
\ena
 \end{e_prop}

 {\it Proof:} See Appendix~\ref{TangPropProof}.
\QED

\begin{figure}[t]
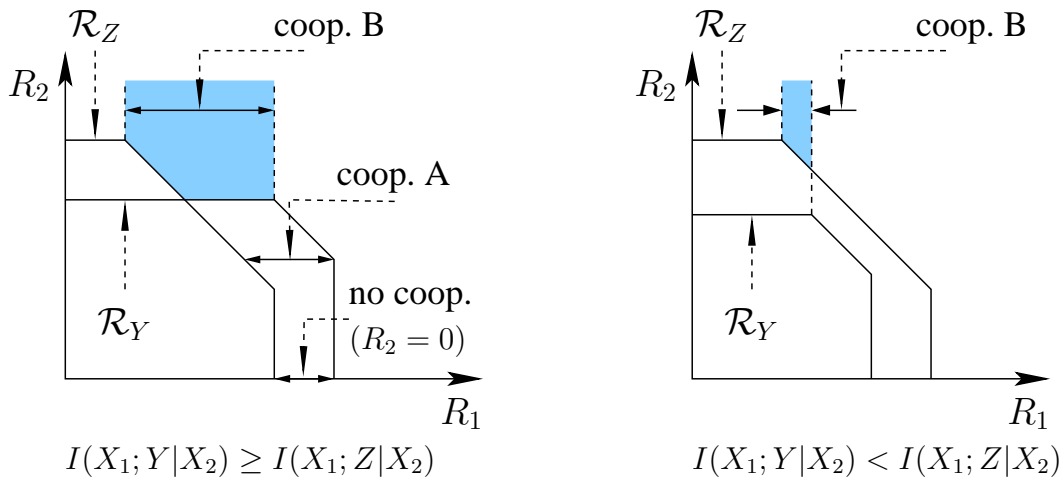

\begin{center}
\input regions_cd_1column.pstex_t
\caption{Pictorial representation for the equivocation gain when cooperation is used for the case (\rn{2}) in Proposition~\ref{TangProp} for the situations $I(X_1;Y|X_2)\ge I(X_1;Z|X_2)$ (left) and $I(X_1;Y|X_2)\ge I(X_1;Z|X_2)$ (right). Pentagons $\mathcal{R}_Y$ and $\mathcal{R}_Z$ express an achievable region for the receiver's MAC and the eavesdropper's MAC, respectively. The cooperation scheme that achieves $\mathcal{C}_A(P^*)$ is labeled ``coop. A'' and the cooperation scheme that achieves $\mathcal{C}_B(P^*)$, is labeled ``coop. B''.}
\label{fig:region_CB2}
\end{center}
\end{figure}

We illustrate both cases, in which $\mathcal{C}_B (P^*)$ is effective, in Figs.~\ref{fig:region_CB1} and~\ref{fig:region_CB2}.
For illustrative purpose, we consider rate-equivocation regions given by $P_{X_1|Q_1}P_{X_2}$. The actual region is obtained by prefixing $P_{X_1|U_1}P_{X_2|U_2}$ as discussed in Appendix~\ref{Append:achievability}. Fig.~\ref{fig:region_CB1} describes the case that satisfies (\ref{eq:effective_condition1}) in the following two situations: $I(X_1;Y|X_2)\ge I(X_1;Z|X_2)$ (left) and $I(X_1;Y|X_2)\ge I(X_1;Z|X_2)$ (right). Fig.~\ref{fig:region_CB2} describes the case that satisfies that satisfies (\ref{eq:effective_condition2}) in the following two situations: $I(X_1;Y|X_2)\ge I(X_1;Z|X_2)$ (left) and $I(X_1;Y|X_2)\ge I(X_1;Z|X_2)$ (right). In the figures ``coop. A'' denotes the cooperation scheme that achieves $\mathcal{C}_A(P^*)$, while ``coop. B'' the cooperation scheme that achieves $\mathcal{C}_B(P^*)$. Observe that in the right subfigures of both figures only cooperation scheme B gives positive equivocation, implying the usefulness of this cooperation scheme.

\section{A Note on the Broadcast Channel with Confidential Messages and a Helping Interferer} \label{sect:BCC_Helper}

Since the effect of $Q_1$ is not completely clear, one might doubt the true effect of $Q_1$. In this section, we discuss about the role of the introduced $Q_1$ by comparing relationship between the broadcast channel with confidential messages (BCC) and the wire-tap channel with a helping interferer. We consider the following two items:
 \begin{itemize}
\item[(1)] The constraint on $R_1$ in the new achievable rate-equivocation region $\mathcal{C}$ involves the term 
\[\min\{I(Q_1;Y|U_2), I(Q_1;Z|U_2)\},\] whereas the capacity equivocation region for the ordinary wire-tap channel does not (c.f., (\ref{eq:wiretap_CE_region})).
\item[(2)] Although by introducing another auxiliary random variable $Q_1$ we have a wider rate-equivocation region, this random variable gives no impact in terms of perfect-secrecy (i.e., (\ref{eq:achievable_PS_rate})).
\end{itemize}
Csisz\'ar and K\"orner show in \cite{Csiszar-Korner78} that for the broadcast channel with confidential messages (BCC), the use of $Q_1$ is essential. 

\begin{figure}[t]
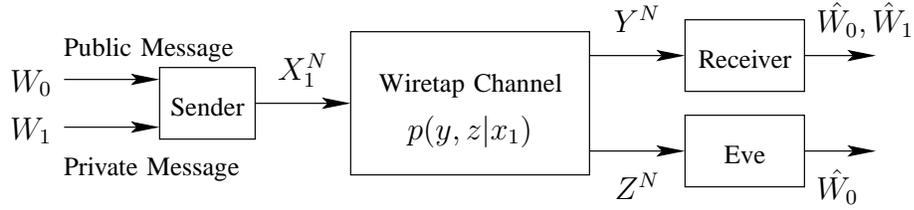

\begin{center}
\input BCC_1column.pstex_t
\caption{The broadcast channel with confidential messages (BCC).} \label{fig:BCC}
\end{center}
\end{figure}

\begin{figure}[t]
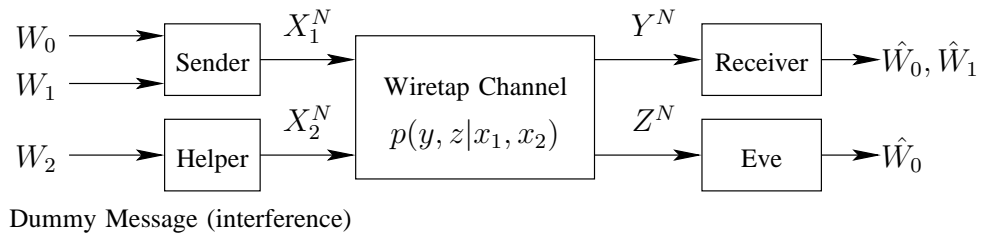

\begin{center}
\input BCChelper_1column.pstex_t
\caption{BC with confidential messages and with a helper.} \label{fig:BCC_helper}
\end{center}
\end{figure}

In the model of BCC (Fig.~\ref{fig:BCC}), there are two receivers, and the sender wishes to send public messages $\mathcal{W}_0$ of rate $R_0$ to both receivers while public messages $\mathcal{W}_1$ of rate $R_1$ is confidential to receiver 2. It is known that the following region is the capacity-equivocation region for the BCC \cite[Theorem 1]{Csiszar-Korner78}
\begin{eqnarray}
\mathcal{C}_{BCC} =  \Big\{ (R_1, R_{\rm e}, R_0): &&0 \le R_0, 0 \le R_{\rm e} \le R_1, \nonumber \\
&&R_0 + R_1 \le I(U_1;Y|Q_1) + \min\{ I(Q_1;Y), I(Q_1;Z)\}, \nonumber \\
&&R_{\rm e} \le  I(U_1;Y|Q_1) - I(U_1;Z|Q_1),  \nonumber \\
&& R_0 \le \min\{ I(Q_1;Y), I(Q_1;Z)\} \Big\} \label{eq:BCC_capacity_region}
\end{eqnarray}
where the random variables satisfy
\begin{eqnarray}
Q_1 \rightarrow U_1 \rightarrow X_1 \rightarrow (Y,Z). \label{eq:Markov_chain3}
\end{eqnarray}
From (\ref{eq:BCC_capacity_region}), the constraint on $R_0 + R_1$ also involves the term $\min \{ I(Q_1;Y), I(Q_1;Z)\}$ (c.f., above item 1). Furthermore, the random variable $Q_1$ is essentially necessary because receiver 2 should estimate $W_0$ from $Z^N$ reliably. Having this in mind, we can argue the \emph{BCC with a helping interferer} as in Fig.~\ref{fig:BCC_helper}, and we can achieve the following rate-equivocation region, that is the convex hull of
\begin{equation}
\mathcal{C}_{BCCH} = \bigcup_{\pi^*} \big\{ \mathcal{C}_A' (\pi^*) \cup \mathcal{C}_B' (\pi^*) \big\}. \label{eq:union_region6}
\end{equation}
where
\begin{eqnarray}
\mathcal{C}_A'(\pi^*) \triangleq  \Big\{ (R_1, R_{\rm e}, R_0): &&0 \le R_0, 0 \le R_{\rm e} \le R_1, \nonumber \\
&&  R_0+ R_1 \le I(U_1;Y|U_2Q_1) + \min\{ I(Q_1;Y|U_2), I(Q_1;Z|U_2)\}, \nonumber \\
&& R_{\rm e} \le I(U_1;Y|U_2Q_1) - I(U_1;Z|U_2Q_1), \nonumber \\
&&  R_0 \le \min\{ I(Q_1;Y|U_2), I(Q_1;Z|U_2)\} \Big\},\n
\label{eq:achievable_region7}
\end{eqnarray}
and
\begin{eqnarray}
\mathcal{C}'_B(\pi^*) =  \Big\{ (R_1, R_{\rm e}, R_0): && 0 \le R_0, 0 \le R_{\rm e} \le R_1, \nonumber \\
&& R_0+ R_1 \le I(U_1;Y|Q_1) + \min\{ I(Q_1;Y), I(Q_1;Z)\}, \nonumber \\
&& R_{\rm e} \le  I(U_1;Y|Q_1) - I(U_1;Z|Q_1), \nonumber \\
&&  R_0 \le \min\{ I(Q_1;Y), I(Q_1;Z)\} \Big\}. \label{eq:achievable_region7b}\n
\end{eqnarray}
Here, $Q_1$, $U_1$, and $U_2$ are auxiliary random variables satisfying the Markov chain conditions (\ref{eq:Markov_chain1}).
Therefore, we think that in the case of the BCC with a helper, the use of $Q_1$ is also \emph{essential}.
Furthermore, when $R_2=0$, the above region reduces to the capacity-equivocation region for the BCC.

\begin{e_rema}
We cannot directly use the achievability scheme from Appendix~\ref{Append:achievability}, and the constraint on $R_{\rm e} \in \mathcal{C}_A'(P^*)$, which is always achieved with $R_2=0$, is smaller than that in $\mathcal{C}_A(P^*)$ given in (\ref{eq:achievable_region5}) for the wire-tap channel with a helper. The main reason for this is that, by setting \[R_2 \le \min\{I(U_2;Y|Q_1), I(U_2;Z|U_1)\}\] as in Appendix~\ref{Append:achievability}, then receiver 2 cannot always decode $W_2$ (and equivalently, $U_2^N$) correctly, and it cannot decode $W_0$ accordingly for some \[R_0 \le \min\{I(Q_1;Y|U_2), I(Q_1;Z|U_2)\}.\] 
{\textcolor{black}{If $I(Q_1;Y|U_2) \le I(Q_1;Z)$ is satisfied, then there is a possibility to have an advantage.}}
On the other hand, in the case of the wire-tap channel with a helper, $W_2$ needs not be decoded by the eavesdropper, so this problem does not occur.
\end{e_rema}

Despite the above remark, from (\ref{eq:BCC_capacity_region}) and (\ref{eq:union_region6}), the cooperation by a helper gives a larger rate-equivocation region compared with the case of the ordinary BCC (with no helpers).
This indicates that the cooperation has an effect even for the BCC case, and observation by Tang et al.~\cite{TLPP2008} is also useful.


\section{Conclusion} \label{sect:conclusion}

We have derived a new achievable rate-equivocation region for a class of wire-tap channels with a helping interferer, which has been shown to be larger than the rate-equivocation region given by \cite{Lai-ElGamal2008}. Our result can naturally adopt the observation given by \cite{TLPP2008}, yielding an even larger rate-equivocation region than the previously known regions. We also discussed about some relationship of our result with the capacity-equivocation over the broadcast channel with confidential messages in order to explain the role of the newly introduced random variable. 




%

\ifCLASSOPTIONcaptionsoff
  \newpage
\fi

\appendices

\section{Achievability of the New Region} \label{Append:achievability}
We shall show an achievability scheme for the region $\mathcal{C}$ via random coding. As in the wire-tap channel \cite{Csiszar-Korner78}, we introduce rate splitting of $R_1$ into $R_{10}$ and $R_{11}$, where $R_{10}$ denotes the rate of messages that can be decoded by both the receiver and the eavesdropper, and $R_{11}$ denotes the rate of messages that can be decoded only by the receiver. First we define the following region:
\begin{eqnarray}
\mathcal{R}_1 = \bigcup_{P_{X_1\textcolor{black}{Q_1}} P_{X_2} P_{YZ|X_1X_2}} \Big\{ (R_1, R_{\rm e}): && R_1 = R_{10} + R_{11}, 0\le R_{10}, 0\le R_{\rm e} \le R_1,\nonumber \\
&& R_{10} \le \min \{I (\textcolor{black}{Q_1};Y|X_2), I(\textcolor{black}{Q_1};Z|X_2) \},  \nonumber \\
&&R_{11} \le I(X_1;Y|X_2\textcolor{black}{Q_1}), \nonumber \\
&& R_{\rm e} \le \max \Big\{ I(X_1;Y|U_2\textcolor{black}{Q_1}) - I(X_1;Z|X_2\textcolor{black}{Q_1}), \n\\
\hadjust{22mm} I(X_1;Y|X_2\textcolor{black}{Q_1}) - I(X_1X_2;Z|\textcolor{black}{Q_1}) \n\\
\hadjust{22mm}+ \min\{ I(X_2;Y|\textcolor{black}{Q_1}), I(X_2;Z|X_1) \} \Big\}. \label{eq:achievable_region2d}
\end{eqnarray}
As discussed in \cite{Csiszar-Korner78}, if $\mathcal{R}_1 $ is achievable, the following region $\mathcal{R}_2$ is achievable by prefixing a conditional PMF  $P_{X_1|U_1}P_{X_2|U_2} $:
\begin{eqnarray}
\mathcal{R}_2 = \bigcup_{P_{\textcolor{black}{Q_1}}  P_{U_1|\textcolor{black}{Q_1}}P_{U_2} P_{X_1|U_1} P_{X_2|U_2} P_{YZ|X_1X_2}} \Big\{ (R_1, R_{\rm e}): && R_1 = R_{10} + R_{11}, 0 \le R_{10}, 0\le R_{\rm e} \le R_1, \nonumber \\
&& R_{10} \le \min \{I (Q_1;Y|U_2), I(Q_1;Z|U_2) \}, \nonumber \\
&& R_{11} \le I(U_1;Y|U_2Q_1), \nonumber \\
&&  R_{\rm e} \le \max \Big\{ I(U_1;Y|U_2\textcolor{black}{Q_1}) - I(U_1;Z|U_2\textcolor{black}{Q_1}), \nonumber \\
\hadjust{22mm}  I(U_1;Y|U_2\textcolor{black}{Q_1}) - I(U_1U_2;Z|\textcolor{black}{Q_1}) \n\\
\hadjust{22mm}+ \min\{ I(U_2;Y|\textcolor{black}{Q_1}), I(U_2;Z|U_1) \} \Big\}. \n\label{eq:achievable_region2b}
\end{eqnarray}
By using the relation $R_{10} = R_1 - R_{11}$, Fourier-Motzkin elimination yields the region $\mathcal{C}$ given by (\ref{eq:achievable_region2}).
Hence, in terms of the achievability to the region $\mathcal{C}$, it suffices to show that the rate-equivocation region $\mathcal{R}_1$  is achievable for every given $P_{Q_1X_1} P_{X_2}$.


Next we show the achievability of $\mathcal{R}_1$, given by (\ref{eq:achievable_region2d}), via random coding and the joint \emph {asymptotic equipartition property} (AEP) \cite{Cover-Thomas91}. We fix a joint PMF $P_{Q_1X_1X_2}^* \triangleq P_{Q_1X_1} P_{X_2}$, and let the target region be denoted by $\mathcal{R}_1(P_{Q_1X_1X_2}^*)$. We consider two cases that will be called Case 1 and Case 2.

\subsection{Case 1: $I(X_1;Y|X_2Q_1) \le I(X_1;Z|X_2Q_1)$}
 In this case, we need to consider only the case $I(X_2;Y|Q_1) \ge I(X_2;Z|Q_1)$, since otherwise the rate-equivocation becomes zero because the first constraint on $R_{\rm e}$ in (\ref{eq:achievable_region2d}) is apparently negative (i.e., it gives a trivial upper-bound on $R_{\rm e}$).
Then, the constraint on $R_{\rm e}$ is expressed as
\begin{eqnarray}
R_{\rm e} \le [I(X_1;Y|X_2Q_1) + \min\{I(X_2;Y|Q_1), I(X_2;Z|X_1)\} -I(X_1X_2;Z|Q_1)]^+. \label{eq:equivocation1}
\end{eqnarray}

\subsubsection{Codebook generation} For a given $P_{Q_1X_1X_2}^* $, we first generate $2^{NR_{10}}$ {\emph independent and identically distributed}(i.i.d.) sequences at random according to \[P_{Q_1^N} (\vect{q}) \triangleq \prod_{n=1}^N P_{Q_1}(q_n),\] and index them as $\vect{q}(i), i \in [1,2^{NR_{10}}]$, with
\begin{equation}
R_{10} \le \min \{I(Q_1;Y|X_2), I(Q_1;Z|X_2)\}. \label{eq:rate_R10a}
\end{equation}
{When $j_1 \le j_2$, $[j_1,j_2]$ denotes the set of all integers from $j_1$ to $j_2$.}
For given $\vect{q}(i), i \in [1,2^{NR_{10}}]$, we generate $2^{NR_{11}}$ i.i.d. sequences at random according to 
\[P_{X_1^N|Q_1^N} (\vect{x}_1|\vect{q}) \triangleq \prod_{n=1}^N P_{X_1|Q_1}(x_{1n}|q_n),\] and index them as $\vect{x}_1(i,b), b \in [1,2^{NR}]$, with
\begin{equation}
\textcolor{black}{R \le I(X_1;Y|X_2Q_1)}. \label{eq:rate_Ra}
\end{equation}
We also generate $2^{NR_2}$ i.i.d. sequences at random according to $P_{X_2^N} (\vect{x}_2) \triangleq \prod_{n=1}^N P_{X_2}(x_{2n})$, and index them as $\vect{x}_2(k), k \in [1,2^{NR_2}]$, with
\begin{equation}
\textcolor{black}{R_2 \le \min\{ I(X_2;Y|Q_1), I(X_2;Z|X_1) \}.} \label{eq:rate_R2a}
\end{equation}
Let 
\begin{equation}
R' \triangleq [R + R_2 - I(X_1X_2;Z|Q_1)]^+ \label{eq:rate_R'a}
\end{equation}
express the rate that exceeds the eavesdropper's ability to decode a sequence reliably. We also define $\mathcal{W} = [1,2^{NR'}]$, $\mathcal{L} = [1,2^{N(R-R')}]$, and $\mathcal{B} = \mathcal{W} \times \mathcal{L} =[1,2^{NR}]$. Note that $R' \le R$ since
\begin{eqnarray}
\hadjust{-5mm}  R -R' \ge I(X_1X_2;Z|Q_1) -I(X_2;Z|\textcolor{black}{X_1}) = \textcolor{black}{I(X_1;Z|Q_1)}.\n
\end{eqnarray}
Hereafter, we assume $R' > 0$ for simplicity. If this is not the case, no security level can be reached, and we achieve only $(R_1,0)$ such that $R_1 \le R$ which is still inside the rate-equivocation region $\mathcal{R}_1$. \textcolor{black}{We call this codebook generation and the encoding and decoding scheme described below \emph{Coding Scheme 1}.}

\subsubsection{Encoding}
 For a given rate-equivocation pair $(R_{10}, R_{11}, R_{\rm e})$ such that $R_1 = R_{10} + R_{11}$ and $R_{\rm e} \le R_1$, we consider the following encoding scheme:
Assume that a secret message $w_1 = (w_{10}, w_{11}) \in \mathcal{W}_1$ with $w_{10} \in \mathcal{W}_{10} \triangleq [1,2^{NR_{10}}]$ and $w_{11} \in \mathcal{W}_{11} \triangleq [1,2^{NR_{11}}]$  is input to sender 1 and a random message $w_2 \in [1, 2^{NR_2}]$ is generated at sender 2.

The encoding function for $w_{11}$ at sender 1 operates in the following stochastic manner:

(\rn{1}) If $R_{11} > R'$, then we divide $\mathcal{W}_{11}$ into $\mathcal{W}$ and $\mathcal{J} \triangleq [1, 2^{N(R_{11} - R')}]$ as $\mathcal{W}_{11} = \mathcal{W} \times \mathcal{J}$.
Let $g$ be the partition that divides $\mathcal{L}$ into $|\mathcal{J}| = 2^{N(R_{11} - R')}$ subsets $\mathcal{L}'_1, \ldots, \mathcal{L}'_{2^{N(R_{11} - R')}}$ with equal cardinalities $2^{N(R - R_{11})}$.
 The encoder determines $(w, l)$ from $w_{11} = (w, j)$ such that $l$ is uniformly chosen from the partition $\mathcal{L}'_{j}$ at random. In this case, there is a one-to-one correspondence between $\{(w, l)\}$ and $[1,2^{NR}]$.

(\rn{2}) If $R_{11} \le R'$, then the encoder obtains $(w, l)$ by setting $w \triangleq w_{11}$ and uniformly choosing $l$ from $\mathcal{L}$ at random. In this case, there is a one-to-one correspondence between $\{(w, l)\}$ and $[1,2^{N(R_{11} + R - R')}]$.

The transmitted sequence from sender 1 is $\vect{x}_1(i, b)$ with $i = w_{10}$ and $b = (w,l) \in [1,2^{NR}]$. Sender 2 transmits the sequence $\vect{x}_2(k)$ with $k = w_2$, where $w_2 \in [1,2^{NR_2}]$ is uniformly selected.

\subsubsection{Decoding:}
 Upon receiving $\vect{y} \in \mathcal{Y}^N $, the receiver seeks a message pair $(\hat{i}, \hat{k})$ such that 
 \[\big(\vect{q}(\hat{i}), \vect{x}_2 (\hat{k}), \vect{y} \big) \in \mathcal{A}_{\epsilon}^{(N)}\] 
 where $\mathcal{A}_{\epsilon}^{(N)}$ denotes the $\epsilon$-jointly typical set \cite{Cover-Thomas91} for any fixed $\epsilon >0$. If there does not exist or there are more than one such sequence, then the receiver declares a decoding error. Then, the receiver seeks a message $\hat{b} = (\hat{w}, \hat{l})$ such that 
 \[\big(\vect{q}(\hat{i}), \vect{x}_1(\hat{i}, \hat{b}), \vect{x}_2(\hat{k}), \vect{y} \big) \in \mathcal{A}_{\epsilon}^{(N)}\] 
 for given $(\hat{i}, \hat{k})$. Having $(\hat{i},\hat{b})$ such that $\hat{b} = (\hat{w}, \hat{l})$, the receiver obtains the estimates of the transmitted message $w_1 = (w_{10}, w_{11})$ by setting
 \eqa
&&\hat{w}_{10} \triangleq \hat{i}, \hat{w}_{11} \triangleq \big(\hat{w}, g(\hat{l}) \big), \quad\mbox{if}\quad R_{11} > R', \quad\mbox{and}\quad\n\\
&&\hat{w}_{10} \triangleq \hat{i}, \hat{w}_{11} \triangleq \hat{w},  \hspace{17mm}\mbox{if}\quad R_{11} \le R'.\n
\ena


\subsubsection{Analysis of Reliability}
The average probability of decoding error for the receiver, denoted by $\overline{P}_{\rm e}^{(N)} (i, b, k) $ provided that $(i, b, k)$ is sent, is upper-bounded as
\begin{equation}
\overline{P}_{\rm e}^{(N)} (i, b, k) \le \overline{P}_{\rm e,1}^{(N)} (i, k) + \overline{P}_{\rm e,2}^{(N)} (b|i,k),
\end{equation}
where $\overline{P}_{\rm e,1}^{(N)} (i, k)$ and $\overline{P}_{\rm e,2}^{(N)} (b|i,k)$ denote the probabilities of decoding error for the first step (estimation of $(i,k)$) and the second step (estimation of $b$ given a true transmitted pair $(i,k)$), respectively.
It is easily seen that the error probability of the first decoding step can be made arbitrarily small for all sufficiently large $N$ by the AEP \cite{Cover-Thomas91} since $R_{10}$ and $R_2$ satisfy (\ref{eq:rate_R10a}), (\ref{eq:rate_R2a}), and
\begin{eqnarray}
R_{10} + R_2 &\le& \min\{ I(Q_1;Y), I(Q_1;Z)  \} + \min\{ I(X_2;Y|Q_1), I(X_2;Z|X_1)\} \n\\&\le& I(Q_1X_2;Y).
\end{eqnarray}
Also, the error probability of the second decoding step can be made arbitrarily small for sufficiently large $N$ by the AEP  and (\ref{eq:rate_Ra}), and so can the probability $\overline{P}_{\rm e}^{(N)} (i, b, k)$.

\subsubsection{Analysis of Equivocation}
The equivocation $R_{\rm e}^{(N)} = \frac{1}{N}H(W_1 | Z^N)$ is lower-bounded by
\begin{eqnarray}
\hadjust{-5mm}   R_{\rm e}^{(N)} =  \frac{1}{N}H(W_{10}W_{11} | Z^N) \nonumber \\
\hadjust{4mm}  \ge  \frac{1}{N}H(W_{10}W_{11} | Z^N W_{10}) \n\\
\hadjust{4mm}=\frac{1}{N}H(W_{11} | Z^N W_{10}),
\end{eqnarray}
where the inequality follows from the fact that conditioning does not increase the entropy. By a similar expansion for $H(W_{11} | Z^N W_{10})$ as in \cite[eq.\ (45)]{Lai-ElGamal2008}, we obtain
\begin{eqnarray}
H(W_{11} | Z^N W_{10}) \ge H(X_1^N X_2^N|W_{10}) -I(X_1^NX_2^N;Z^N|W_{10}) - H(X_1^NX_2^N|W_{10}W_{11}Z^N). \label{eq:lower_bound_equivocation}
\end{eqnarray}
We shall consider bounding each term in (\ref{eq:lower_bound_equivocation}). For the first term, we have
\begin{equation}
H(X_1^N X_2^N|W_{10}) = H(X_1^N|W_{10}) + H(X_2^N)\n\\
\end{equation}
and
\begin{equation}
 H(X_1^N|W_{10})  = H(X_1^N | W_{10}Q_1^N) = H(X_1^N | Q_1^N),\n\\
\end{equation}
where the first equality is due to the fact that $Q_1^N$ is a deterministic function of $W_{10}$, while \textcolor{black}{the last equality follows from the Markov chain relationship $W_{10} \rightarrow Q_1^N \rightarrow X_1^N$}.
Since the codewords are generated according to i.i.d. distributions, it follows that 
\begin{eqnarray}
H(X_1^N | Q_1^N) = NH(X_1|Q_1) \ge NR,
\ena
and
\eqa
H(X_2^N) = NH(X_2) \ge NR_2. \label{eq:entropy1}
\end{eqnarray}
It is sufficient that we directly replace these inequalities with \[NH(X_1|Q_1) \ge NI(X_1;Y|X_2Q_1)\] and \[NH(X_2) \ge N \min\{I(X_2;Y|Q_1), I(X_2;Z|X_1) \}.\] For the second term in (\ref{eq:lower_bound_equivocation}), we expand
\begin{eqnarray}
\hadjust{-7mm} I(X_1^NX_2^N;Z^N|W_{10}) = H(Z^N|W_{10}) - H(Z^N|X_1^NX_2^NW_{10}), \n\\ \label{eq:mutual_info1}
\end{eqnarray}
for which we have
\begin{eqnarray}
\hadjust{-5mm} H(Z^N|W_{10}) = H(Z^N|Q_1^N) = NH(Z|Q_1) \label{eq:entropy2}
\end{eqnarray}
due to the fact that $Q_1^N$ is a deterministic function of $W_{10}$, the Markov chain relationship  $W_{10} \rightarrow Q_1^N \rightarrow Z^N$, and an i.i.d. distribution for $Z^N$ given $Q_1^N$. We also have
\begin{eqnarray}
H(Z^N|X_1^NX_2^NW_{10}) &=& H(Z^N|X_1^NX_2^NQ_1^N) = NH(Z|X_1X_2Q_1). \label{eq:entropy3}
\end{eqnarray}
It follows from (\ref{eq:entropy2}) and (\ref{eq:entropy3}) that (\ref{eq:mutual_info1}) becomes
\begin{eqnarray}
\hadjust{-5mm} I(X_1^NX_2^N;Z^N|W_{10}) = NI(X_1X_2;Z|Q_1). \label{eq:mutual_info3}
\end{eqnarray}
We now consider the third term in (\ref{eq:lower_bound_equivocation}).
Consider decoding of $l$ given $w_{1} = (w_{10}, w_{11}) \in \mathcal{W}_1$ by observing $\vect{z} \in \mathcal{Z}^N$.
For the case $R_{11} > R'$, since this decoder knows $j \in \mathcal{J}$, which is  given by $w_{11} =(w,j) \in \mathcal{W}_{11}$, and using the following inequalities
\begin{equation}
 \frac{1}{N}\log_2 |\mathcal{L}'_{j} | + R_2 \le R - R' + R_2 = I(X_1X_2;Z|Q_1),\n
\end{equation}
and 
\begin{equation}
 \frac{1}{N}\log_2 |\mathcal{L}'_{j} | \le R \le I(X_1;Z|X_2Q_1),  \label{eq:rate_L'}
\end{equation}
the average probability of decoding error can be made arbitrarily small for sufficiently large $N$. Note that, in this case, 
\[R \le I(X_1;Y|X_2Q_1) \le I(X_1;Z|X_2Q_1).\] 
For the case $R_{11} \le R'$, we also have
\begin{equation}
 \frac{1}{N}\log_2 |\mathcal{L} | + R_2 \le R - R' + R_2 = I(X_1X_2;Z|Q_1),\n
\end{equation}
and
\begin{equation}
 \frac{1}{N}\log_2 |\mathcal{L} | \le R \le I(X_1;Z|X_2Q_1).  \label{eq:rate_L}
\end{equation}
Again, the average probability of decoding error can be made arbitrarily small with all sufficiently large $N$.
Therefore, by Fano's inequality \cite{Cover-Thomas91}, for any given $\epsilon' > 0$, we have
\begin{equation}
\frac{1}{N} H(X_1^NX_2^N|W_{10}W_{11}Z^N) \le \epsilon' \label{eq:third_term}
\end{equation}
 for sufficiently large $N$. Substituting (\ref{eq:entropy1}), (\ref{eq:mutual_info3}), and (\ref{eq:third_term}) into (\ref{eq:lower_bound_equivocation}) yields, for any given $\epsilon' > 0$,
\begin{equation}
R_{\rm e}^{(N)}  \ge R + R_2 - I(X_1X_2;Z|Q_1) - \epsilon' \label{eq:equivocation2}\n
\end{equation}
 for $N$ sufficiently large. Since we can choose any pair of $R$ and $R_2$ subject to (\ref{eq:rate_Ra}) and (\ref{eq:rate_R2a}), there exist $R$ and $R_2$ such that, for any $\epsilon' >0$,
\begin{eqnarray}
R_{\rm e}^{(N)}  \ge I(X_1;Y|X_2Q_1) + \min\{ I(X_2;Y|Q_1), I(X_2;Z|X_1)\}- I(X_1X_2;Z|Q_1) - \epsilon'. \label{eq:equivocation3}
\end{eqnarray}
Hence, it follows from (\ref{eq:rate_R'a}) and (\ref{eq:equivocation3}) that any equivocation $R_{\rm e}$ satisfying (\ref{eq:equivocation1}) is achievable.

\subsection{Case 2: $I(X_1;Y|X_2Q_1) > I(X_1;Z|X_2Q_1)$}
In this case, if $I(X_2;Y|Q_1) \ge I(X_2;Z|Q_1)$, then the constraint on $R_{\rm e}$ is given by (\ref{eq:equivocation1}). We can use Coding Scheme 1 discussed in Case 1 with a slight modification. We set
\begin{eqnarray}
R' = \big[R &+& \min\{ I(X_2;Y|Q_1), I(X_2;Z|X_1) \} - I(X_1X_2;Z|Q_1)\big]^+, \label{eq:rate_R'b}\n
\end{eqnarray}
and we assume that $R' >0$, because no security level is obtained otherwise. Then, for the analysis of equivocation, the left hand side of (\ref{eq:rate_L'}) is bounded as
\begin{eqnarray}
\frac{1}{N} \log_2 |\mathcal{L}'_j| \le R - R' &=& I(X_1X_2;Z|Q_1) - \min\{I(X_2;Y|Q_1), I(X_2;Z|X_1)\}. \n
\end{eqnarray}
Since $I(X_2;Z|Q_1) \le \min\{I(X_2;Y|Q_1), I(X_2;Z|X_1)\}$ and
\begin{eqnarray}
I(X_2;Z|Q_1) &=& H(X_2) - H(X_2|Q_1Z)\n\\  &\le& H(X_2) - H(X_2|Q_1X_1Z) \n\\&=&I(X_2;Z|X_1)\n
\end{eqnarray}
where the last equality follows from the Markov chain relationship \[Q_1 \rightarrow (X_1,Z) \rightarrow X_2,\] we have
(\ref{eq:rate_L'}) if $R_{11} > R'$. From the same reasoning, we also have (\ref{eq:rate_L}) if $R_{11} \le R'$.
Other arguments are quite similar to those for Case 1, and we can show that any rate-equivocation pair $(R_1, R_{\rm}) \in \mathcal{R}_1 (P_{Q_1X_1X_2}^*)$ is achievable.

We then consider the case $I(X_2;Y|Q_1) \le I(X_2;Z|Q_1)$.
In this case, the constraint on $R_{\rm e}$ in (\ref{eq:achievable_region2d}) is given by
\begin{equation}
R_{\rm e} \le I(X_1;Y|X_2Q_1) - I(X_1;Z|X_2Q_1),\n
\end{equation}
which can be achieved by a similar coding/decoding scheme to Coding Scheme 1 by letting 
\eqa
R &\le& I(X_1;Y|X_2Q_1),\mbox{and}\n\\
R' &=& [R_1 - I(X_1;Z|X_2Q_1)]^+.\n
\ena
In this case, $R_2$ can be arbitrarily set in the range $0 \le R_2 \le I(X_2;Y|Q_1)$.
\textcolor{black}{We call this coding scheme \emph{Coding Scheme 2}.}

To show the equivocation at the eavesdropper, note that
\eqa
R_{\rm e}^{(N)} &=& \frac{1}{N}H(W_{10}W_{11}|Z^N) \n\\
&\ge& \frac{1}{N} H(W_{11} | Z^N X_2^N W_{10} ).\n
\ena
Similarly to the derivation \cite[eq.\ (49)]{Lai-ElGamal2008}, we obtain
\begin{eqnarray}
H(W_{11} | Z^N X_2^N W_{10} ) &\ge& H(X_1^N|W_{10}) -I(X_1^N;Z^N|W_{10}X_2^N) - H(X_1^N|W_{10}W_{11}Z^NX_2^N),\n
\end{eqnarray}
in which the right hand side is lower-bounded by 
\[N(I(X_1;Y|X_2Q_1) - I(X_1;Z|X_2Q_1) -\epsilon),\] 
 for any given $\epsilon > 0$, for all sufficiently large $N$. This completes the proof of the achievability to the region $\mathcal{R}_1(P_{Q_1X_1X_2}^*)$.

\section{An Achievable Scheme for the Region $\mathcal{R}_B$} \label{append:RB}




We give an achievability scheme for the region $\tilde{\mathcal{C}}$ given in (\ref{eq:union_region2}). Let $\pi^*=P_{Q_1U_1U_2X_1X_2}^*$ and $\mathcal{R}_B(\pi^*)$ be defined as
\begin{eqnarray}
\mathcal{R}_B(\pi^*) =  \Big\{ (R_1, R_{\rm e}): && R_1 = R_{10}+R_{11}, 0 \le R_{10}, \n\\
&& 0 \le R_{\rm e} \le R_1, \nonumber \\
&&  R_{10} \le \min\{I(Q_1;Y) , I(Q_1;Z)\}, \nonumber \\
&&  R_{11} \le I(U_1;Y|Q_1), \nonumber \\
&& R_{\rm e} \le  I(U_1;Y|Q_1) - I(U_1;Z|Q_1) \Big\}. \label{eq:achievable_region6}
\end{eqnarray}
By virtue of Fourier-Motzkin elimination, it is readily shown that
\begin{equation}
\bigcup_{\pi^*} \mathcal{C}_B(\pi^*) = \bigcup_{\pi^*} \mathcal{R}_B(\pi^*),
\end{equation}
and from (\ref{eq:union_region1}) and (\ref{eq:union_region2}),
\begin{eqnarray}
\tilde{\mathcal{C}} &=& \bigcup_{\pi^*} \big\{ \mathcal{C}_A (\pi^*) \cup \mathcal{R}_B(\pi^*) \big\} \nonumber \\
&=& \mathcal{C} \cup  \bigcup_{\pi^*} \mathcal{R}_B(\pi^*).\n
\end{eqnarray}

The region $\mathcal{C}$ is achievable by the coding method given in Section~\ref{sect:new_RE_region}. Therefore, if we have an achievability scheme to achieve $\mathcal{R}_B(P_{Q_1U_1U_2X_1X_2}^*)$ for any given $P_{Q_1U_1U_2X_1X_2}^* \in \mathcal{P}^*$, then the region $\tilde{\mathcal{C}}$ is also achievable.

We turn to showing an achievable scheme to the region $\mathcal{R}_B(P_{Q_1U_1U_2X_1X_2}^*)$ for arbitrarily fixed $\pi^*=P_{Q_1U_1U_2X_1X_2}^* \in \mathcal{P}^*$. The description of a achievability scheme is a combination of the scheme in Appendix~\ref{Append:achievability} and the scheme given in \cite{TLPP2008}.

\section{Proof of Proposition~\ref{TangProp}}
\label{TangPropProof}

 The condition $I(U_1;Y|Q) > I(U_1;Z|Q) $ is necessary since otherwise there is no equivocation in $\mathcal{C}_B(P^*)$. As we have seen in (\ref{eq:another_expression}), there are three cases for which the region $\mathcal{C}_A(P^*)$ is of different form.

If $I(U_2;Y|Q_1) \le I(U_2;Z|Q_1)$, then the constraint on $R_{\rm e} \in \mathcal{C}_A(P^*)$ is given by 
\[R_{\rm e} \le [I(U_1;Y|U_2Q_1) - I(U_1;Z|U_2Q_1)]^+.\] 
Then the constraint on $R_{\rm e} \in \mathcal{C}_B(P^*)$ has an effect iff
\begin{equation}
\hspace{-1mm} I(U_1;Y|Q_1) - I(U_1;Z|Q_1) \ge I(U_1;Y|U_2Q_1) - I(U_1;Z|U_2Q_1). \label{eq:condition_case1}
\end{equation}
First note that
\begin{eqnarray}
\hadjust{0mm}I(U_1;Y|Q_1) - I(U_1;Z|Q_1) - (I(U_1;Y|U_2Q_1) - I(U_1;Z|U_2Q_1)) \nonumber \\
\hadjust{22mm} = I(U_1;Z|U_2Q_1) - I(U_1;Z|Q_1) - (I(U_1;Y|U_2Q_1) - I(U_1;Y|Q_1)). \label{eq:difference1}
\end{eqnarray}
Since
\begin{eqnarray}
I(U_1;Z|U_2Q_1) - I(U_1;Z|Q_1) &=&  I(U_1U_2;Z|Q_1)- I(U_1;Z|Q_1) - I(U_2;Z|Q_1) \nonumber \\
&=&  I(U_2;Z|U_1)- I(U_2;Z|Q_1)\n
\end{eqnarray}
and also 
\begin{eqnarray}
\hadjust{-7mm} I(U_1;Y|U_2Q_1) - I(U_1;Y|Q_1) =   I(U_2;Y|U_1)- I(U_2;Y|Q_1),\n
\end{eqnarray}
then (\ref{eq:difference1}) becomes
\begin{eqnarray}
\hadjust{0mm}  I(U_1;Y|Q_1) - I(U_1;Z|Q_1) - (I(U_1;Y|U_2Q_1) - I(U_1;Z|U_2Q_1)) \nonumber \\
\hadjust{22mm}   = I(U_2;Z|U_1)- I(U_2;Z|Q_1) - (I(U_2;Y|U_1)- I(U_2;Y|Q_1))). \label{eq:difference1b}
\end{eqnarray}
Therefore, (\ref{eq:condition_case1}) holds iff
\begin{equation}
 I(U_2;Z|U_1)- I(U_2;Z|Q_1)  \ge I(U_2;Y|U_1) - I(U_2;Y|Q_1), \label{eq:condition_case1b}\n
\end{equation}
leading to (\ref{eq:effective_condition1}).

If $I(U_2;Z|Q_1) \le I(U_2;Y|Q_1) \le I(U_2;Z|U_1)$, then the constraint on $R_{\rm e}\in\mathcal{C}_A(P^*)$ is given by 
\[R_{\rm e} \le  [I(U_1U_2;Y|Q_1) - I(U_1U_2;Z|Q_1)]^+.\] Then the constraint on $R_{\rm e} \in \mathcal{C}_B(P^*)$ has an effect iff
\begin{eqnarray}
I(U_1;Y|Q_1) - I(U_1;Z|Q_1) &\ge& I(U_1U_2;Y|Q_1) - I(U_1U_2;Z|Q_1). \label{eq:condition_case2}
\end{eqnarray}
We note that
\begin{eqnarray}
\hadjust{0mm}I(U_1;Y|Q_1) - I(U_1;Z|Q_1) - (I(U_1U_2;Y|Q_1) - I(U_1U_2;Z|Q_1)) \nonumber \\
\hadjust{22mm}   = I(U_2;Z|U_1) - I(U_2;Y|U_1). \label{eq:difference2}
\end{eqnarray}
Therefore, (\ref{eq:condition_case2}) holds iff (\ref{eq:effective_condition2}) holds.

If $I(U_2;Z|U_1)  \le I(U_2;Y|Q_1) $, then the constraint on $R_{\rm e}\in\mathcal{C}_A(P^*)$ is given by $R_{\rm e} \le [I(U_1U_2;Y|Q_1) - I(U_1;Z|Q_1)]^+ $. In this case, since it always holds
\begin{eqnarray}
I(U_1;Y|Q_1) - I(U_1;Z|Q_1) &\le& I(U_1U_2;Y|Q_1) - I(U_1;Z|Q_1),
\end{eqnarray}
the constraint on $R_{\rm e} \in \mathcal{C}_B(P^*)$ has \emph{no effect}.
\QED

\section{The Wire-Tap Channel with a Deaf-Interferer}
\label{sec: deaf-helper}

In wireless network settings, sender 2 (the helper) in the wire-tap channel with a helper can observe a noisy sequence of the transmitted sequence $X_1^N$ from sender 1.
Let $Y_1^N$ denote the sequence observed by sender 2.
For some security systems, it is desired to avoid leaking information about $W_1$ to sender 2, which motivates the introduction of another type of the wire-tap channel with a helper, called the wire-tap channel with a \emph{deaf-helper} (a \emph{deaf-interferer}) \cite{Lai-ElGamal2008}.

The wire-tap channel with a deaf-helper looks like the relay-eavesdropper channel, in which a relay node observes $Y_1^N$ and helps to increase the rate of $W_1$ or the equivocation at the eavesdropper.
Note that in this channel model, the relay node might (partially) decode the  message $W_1$ for the cooperation.
On the other hand, the scenario of the wire-tap channel with a deaf-helper describes the setting in which sender 1 with secret messages does not fully trust the other sender (the helper) but still wishes to get help from the user cooperation.
As in \cite{Lai-ElGamal2008}, we assume that sender 2 is not malicious, and willing to help the communication from sender 1 to the receiver. {Since sender 2 ''forwards'' a dummy sequence instead of forwarding a (partial) message of sender 1, the cooperation scheme is called a \emph{noise-forwarding (NF)} strategy.}

In this setting, a rate-equivocation region is defined by introducing an additional security constraint as follows:
\begin{e_defin}
A \emph{rate-equivocation pair} $(R_1, R_{\rm e})$ is said to be \emph{achievable} if there exists a sequence of $(N,M_1)$ codes such that for every $\epsilon > 0$,
\begin{eqnarray}
R_1  &\ge&  \frac{\log_2 M_1}{N} - \epsilon,  \n\\
P_{\rm e}^{(N)}  &\le&  \epsilon,  \n\\
R_{\rm e}^{(N)} &\triangleq& \frac{1}{N}H(W_1|Z^N)  \ge  R_{\rm e} -\epsilon,\n\\
R_{\rm s}^{(N)} &\triangleq& \frac{1}{N}H(W_1|Y_1^N X_2^N)  \ge  R_{\rm e} -\epsilon\n
\end{eqnarray}
for all sufficiently large $N$.
\end{e_defin}

We conjecture that the convex hull of the following rate-equivocation region is achievable 
\begin{eqnarray}
\hadjust{0mm}\mathcal{C}_{\rm DH} = \bigcup_{P_{\textcolor{black}{Q_1}}  P_{U_1|\textcolor{black}{Q_1}}P_{U_2} P_{X_1|U_1} P_{X_2|U_2} P_{YZ|X_1X_2}} \Big\{ (R_1, R_{\rm e}): 0 \le R_{\rm e} \le R_1, \nonumber \\
\hadjust{40mm}R_1 \le I(U_1;Y|U_2\textcolor{black}{Q_1}) + \min\{ I(\textcolor{black}{Q_1};Y|U_2), I(\textcolor{black}{Q_1};Z|U_2)\}, \nonumber \\
\hadjust{40mm}R_{\rm e} \le \max \Big\{ R_3' - I(U_1;Z|U_2\textcolor{black}{Q_1}) -I(U_2;Y|\textcolor{black}{Q_1}), R_3' - I(U_1U_2;Z|\textcolor{black}{Q_1}) \Big\} \nonumber \\
\hadjust{40mm}R_{\rm e} \le [I(U_1;Y|U_2\textcolor{black}{Q_1}) - I(U_1;Y_1|U_2\textcolor{black}{Q_1})]^+ \Big\} \label{eq:achievable_region4}\n
\end{eqnarray}
where \[R_2' = \min\{I(U_2;Y|\textcolor{black}{Q_1}), I(U_2,Z|U_1) \},\] \[R_3'=I(U_1;Y|U_2\textcolor{black}{Q_1})+R_2',\] and $Q_1$, $U_1$ and $U_2$ are auxiliary random variables satisfying the Markov chain conditions given by (\ref{eq:Markov_chain1}).

As for the perfect-secrecy rate, the above achievable rate-equivocation region reduces to the following result, which is the same as that given in \cite[Theorem 6]{Lai-ElGamal2008}.
\begin{e_theo}
The perfect-secrecy rate for the wire-tap channel with a deaf-helper, given by
\begin{eqnarray}
R_1 &=& \sup_{P_{U_1X_1}P_{U_2X_2}} \min \{ R_{\mbox{e},1}, R_{\mbox{e},2}\},\n
\end{eqnarray}
where
\begin{eqnarray}
R_{\mbox{e},1} &\triangleq& \max \big\{ I(U_1;Y|U_2) - I(U_1;Z|U_2) + R_2' -I(U_2;Y|), I(U_1;Y|U_2) + R_2' - I(U_1U_2;Z)\big\}, \mbox{and} \nonumber \\
R_{\mbox{e},2} &\triangleq& [I(U_1;Y|U_2) + R_2' - I(U_1;Y_1|U_2)]^+, \n\label{eq:perfect-secrecy}
\end{eqnarray}
is achievable.
\end{e_theo}

\bibliographystyle{IEEEtran}
\bibliography{ref_isit11c}

\end{document}